\documentclass[conference,twocolumn,oneside,10pt]{IEEEtran}
\IEEEoverridecommandlockouts

\usepackage{cite}
\ifCLASSINFOpdf
\usepackage[pdftex]{graphicx}
\else
\fi
\usepackage[cmex10]{amsmath}
\usepackage{amssymb,amsbsy,amsfonts,dsfont,amsthm,bm}

\usepackage{algorithm}
\usepackage{algpseudocode}
\usepackage{array}
\usepackage{mdwmath}
\usepackage{mdwtab}
\usepackage[]{caption}
\usepackage[caption=false,font=footnotesize]{subfig}
\usepackage{fixltx2e}
\usepackage{url}
\usepackage{latexsym,supertabular}
\usepackage{tikz}
\usetikzlibrary{shapes,arrows,shadows,automata,backgrounds,chains,matrix,patterns,petri,trees,calc}

\usepackage{balance}
\usepackage{multicol}
\usepackage{multirow}
\usepackage{booktabs}
\usepackage[capitalize]{cleveref}
\crefrangeformat{equation}{Equations~(#3#1#4) to~(#5#2#6)}

\newcommand{\bi}{\begin{itemize}}
\newcommand{\ei}{\end{itemize}}
\newcommand{\be}{\begin{eqnarray}}
\newcommand{\ee}{\end{eqnarray}}

\newcommand{\argmin}[1]{\underset{#1}{\text{arg min}}}
\newcommand{\mx}[1]{\mathbf{\bm{#1}}} % Matrix command.
\newcommand{\vc}[1]{\mathbf{\bm{#1}}} % Vector command.

\usepackage{eqparbox}

\usepackage{paralist}

\usepackage{etoolbox}
\newtoggle{FigPosEndPaper}
\settoggle{FigPosEndPaper}{false} %true or false.
\newbool{EqOneColumn}
\setbool{EqOneColumn}{false} %true or false.
\newbool{HomogeneousWSN}
\setbool{HomogeneousWSN}{true} %true or false.
\ifbool{HomogeneousWSN}
	{\newcommand{\SigmaO}{\sigma_{\text{o}}^2}
	\newcommand{\SigmaC}{\sigma_{\text{c}}^2}}
	{\newcommand{\SigmaO}{\sigma_{\text{o}i}^2}
	\newcommand{\SigmaC}{\sigma_{\text{c}i}^2}}

\usepackage{dblfloatfix}

\usepackage{float}
\usepackage[numbers,sort&compress,square,comma]{natbib}
\newcommand{\AlgParBox}[1]{\hspace{-3pt} \parbox[t]{\dimexpr\linewidth-\algorithmicindent}{#1\strut}}

\begin{document}

% paper title can use linebreaks \\ within to get better formatting as desired.
\title{Limited-Feedback-Based Channel-Aware Power Allocation for Linear Distributed Estimation}

% author names and IEEE memberships. note positions of commas and nonbreaking spaces (~). LaTeX will not break a structure at a ~ so this keeps an author's name from being broken across two lines.
% use \thanks{} to gain access to the first footnote area.
% a separate \thanks must be used for each paragraph as LaTeX2e's \thanks was not built to handle multiple paragraphs.

\author{
	\IEEEauthorblockN{Mohammad~Fanaei, %~\IEEEmembership{Student~Member,~IEEE},
	Matthew~C.~Valenti, %~\IEEEmembership{Senior~Member,~IEEE} \\
	and
	Natalia~A.~Schmid%,~\IEEEmembership{Member,~IEEE}
	}
	\IEEEauthorblockA{Lane Department of Computer Science and Electrical Engineering \\
		West Virginia University, Morgantown, WV, U.S.A. \\
		E-mail: mfanaei@mix.wvu.edu, valenti@ieee.org, and natalia.schmid@mail.wvu.edu.
		%\vspace{-0.28cm}
		}
        % John~Doe,~\IEEEmembership{Fellow,~OSA,}
        % and~Jane~Doe,~\IEEEmembership{Life~Fellow,~IEEE}% <-this % stops a space.
%\thanks{E-mail: mfanaei@mix.wvu.edu, \{matthew.valenti and natalia.schmid\}@mail.wvu.edu.}% <-this % stops a space.
%\thanks{Manuscript received April 19, 2005; revised January 11, 2007.}
\thanks{This work was supported in part by the Office of Naval Research under Award No.~N00014--09--1--1189.
	%1000409W.
	%The contributions of M.~Fanaei and M.C.~Valenti were sponsored in part by the National Science Foundation under Award No.~CNS--0750821.
	The work of M.~Fanaei is sponsored in part by the National Science Foundation under Award~No.~IIA--1317103.
	%1004185R.
	}
}

\maketitle

%---------------------------------------------------------------------------------------------------------------
\begin{abstract}
%\boldmath
This paper investigates the problem of distributed best linear unbiased estimation (BLUE) of a random parameter at the fusion center (FC) of a wireless sensor network (WSN). In particular, the application of limited-feedback strategies for the optimal power allocation in distributed estimation is studied. In order to find the BLUE estimator of the unknown parameter, the FC combines spatially distributed, linearly processed, noisy observations of local sensors received through orthogonal channels corrupted by fading and additive Gaussian noise. Most optimal power-allocation schemes proposed in the literature require the feedback of the exact instantaneous channel state information from the FC to local sensors. This paper proposes a limited-feedback strategy in which the FC designs an optimal codebook containing the optimal power-allocation vectors, in an iterative offline process, based on the generalized Lloyd algorithm with modified distortion functions. Upon observing a realization of the channel vector, the FC finds the closest codeword to its corresponding optimal power-allocation vector and broadcasts the index of the codeword. Each sensor will then transmit its analog observations using its optimal quantized amplification gain. This approach eliminates the requirement for infinite-rate digital feedback links and is scalable, especially in large WSNs.
\end{abstract}
% IEEEtran.cls defaults to using nonbold math in the Abstract. This preserves the distinction between vectors and scalars. However, if the journal you are submitting to favors bold math in the abstract, then you can use LaTeX's standard command \boldmath at the very start of the abstract to achieve this.


% For peer review papers, you can put extra information on the cover page as needed:
% \fi

% For peerreview papers, this IEEEtran command inserts a page break and creates the second title. It will be ignored for other modes.
\IEEEpeerreviewmaketitle

%---------------------------------------------------------------------------------------------------------------
\section{Introduction}
\label{Sec:Intro}
% The very first letter is a 2 line initial drop letter followed by the rest of the first word in caps.
% form to use if you need the single drop letter followed by normal text (unknown if ever used by IEEE):
% You must have at least 2 lines in the paragraph with the drop letter.
%\IEEEPARstart{}{}
Wireless sensor networks (WSNs) are typically formed by spatially distributed sensors with limited communications and processing capabilities that cooperate with each other to achieve a common goal. One of the most important applications of such networks is {\em distributed estimation}, which is a key enabling technology for a wider range of applications such as event classification and object tracking.
% event detection, classification, and object tracking.
In a WSN performing distributed estimation, sensors make noisy observations that are correlated with an unknown parameter to be estimated, process their local observations, and send their processed data to a fusion center (FC), which then combines all of the locally processed samples to perform the ultimate global estimation.

Recently, the problem of distributed estimation in WSNs has extensively been studied in the literature~\cite{Xiao06,Luo05,Xiao08,Banavar10,Cui07Diversity}.
%\cite{Xiao05,Xiao06,Luo05,Xiao08,Cui07Diversity,Banavar10,Chaudhary13,RibeiroGiannakis06a,RibeiroGiannakis06b,Ishwar2005}
The type of local processing that is performed on each sensor's noisy observation before it is transmitted to the FC differentiates these works and can be either a local quantization~\cite{Xiao06,Luo05} or an amplify-and-forward strategy~\cite{Banavar10,Cui07Diversity,Xiao08}.
%One of the main factors that differentiates most of these works is the type of local processing that is performed on each sensor's noisy observation before it is transmitted to the FC. Different design and implementation considerations have been addressed for the case of sending a quantized version of local observations to the FC in~\cite{Xiao05,Xiao06,Luo05,RibeiroGiannakis06a,RibeiroGiannakis06b,Ishwar2005}. The transmission of the analog local observations using an amplify-and-forward strategy has been studied in~\cite{Cui07Diversity,Xiao08,Banavar10,Chaudhary13}.
We consider the second approach in this paper due to its simplicity and practical feasibility. One of the main issues to be addressed in the case of analog amplify-and-forward local processing is finding the optimal local amplification gains.
%~\citep{Banavar10,Cui07Diversity,Xiao08}.
The values of these gains set the instantaneous transmit power of sensors; therefore, we refer to their determination as the {\em power allocation} to sensors.

Cui~et~al.~\cite{Cui07Diversity} have proposed an optimal power-allocation scheme to minimize the variance of the best linear unbiased estimator (BLUE) for a random scalar parameter at the FC of a WSN, given a total transmission-power constraint in the network. In their proposed approach, the optimal local amplification gains depend on the instantaneous fading coefficients of the channels between the sensors and FC. Therefore, in order for the FC to achieve the minimum estimation variance of the BLUE estimator, it must feed the exact channel fading gains back to sensors through infinite-rate, error-free links. This requirement is not practical in most WSN applications, especially when the number of sensors is large. In this paper, we investigate the application of a {\em limited-feedback strategy} for the optimal power-allocation scheme proposed in~\cite{Cui07Diversity}. We use the generalized Lloyd algorithm with modified distortion functions to design an optimal codebook, which is then used to
% optimally
quantize the space of the optimal power-allocation vectors used by the sensors to set their amplification gains.
% Similar approaches have been used to design the optimal codebook for other applications of limited feedback; e.g., the approach proposed by~Liu~and~Chen~\cite{Liu12LimitedFeedbackOFDM} for adaptive power allocation of decode-and-forward relay networks.

Note that the approach proposed in this paper is different from other works that have applied limited feedback to distributed estimation. In particular, the phrase ``limited feedback'' has a different meaning in this paper compared to other works in the field of distributed estimation. For example, Banavar~et~al.~\cite{Banavar10} have investigated the effects of feedback error and the impact of the availability of different amounts of full, partial, or no channel state information at local sensors on the estimation variance of the BLUE estimator of a scalar random parameter at the FC. %\c{S}enol~and~Tepedelenlio\u{g}lu~\cite{Senol08} have studied the problem of distributed estimation using the amplify-and-forward local processing over {\em unknown} parallel fading channels. They have proposed a two-phase approach of pilot-based channel estimation followed by source parameter estimation at the FC. Furthermore, the case in which the estimated channel between each sensor and the FC is available at the sensor for local power optimization has been considered.
The problem of distributed estimation using amplify-and-forward local processing over {\em unknown} parallel fading channels is studied in~\cite{Senol08}. A two-phase approach is proposed based on pilot-based channel estimation followed by source parameter estimation at the FC, where the estimated channel between each sensor and the FC is used at the sensor for local power optimization.

The rest of this paper is organized as follows: In Section~\ref{Sec:SystemModel}, the system model of the WSN under study is introduced.
% and its parameters are introduced.
% Section~\ref{Sec:ProbStatement} summarizes the analysis of the power-allocation scheme proposed in~\cite{Cui07Diversity} and discusses why it is not practical, especially in large-scale WSNs. In Section~\ref{Sec:LimitedFeedbackIntro}, the main concepts of our proposed limited-feedback strategy are introduced.
Section~\ref{Sec:ProbStatement} summarizes the analysis of the power-allocation scheme proposed in~\cite{Cui07Diversity} and introduces the main concepts of our proposed limited-feedback strategy. Details of the implementation of the proposed approach are discussed in Section~\ref{Sec:CodeDesign}. The numerical results are presented in Section~\ref{Sec:NumResults}, and the paper is concluded in Section~\ref{Sec:Conclusions}.

%---------------------------------------------------------------------------------------------------------------
\section{System Model}
\label{Sec:SystemModel}
\begin{figure}[!t]
\centering
\includegraphics[width=0.83\linewidth]{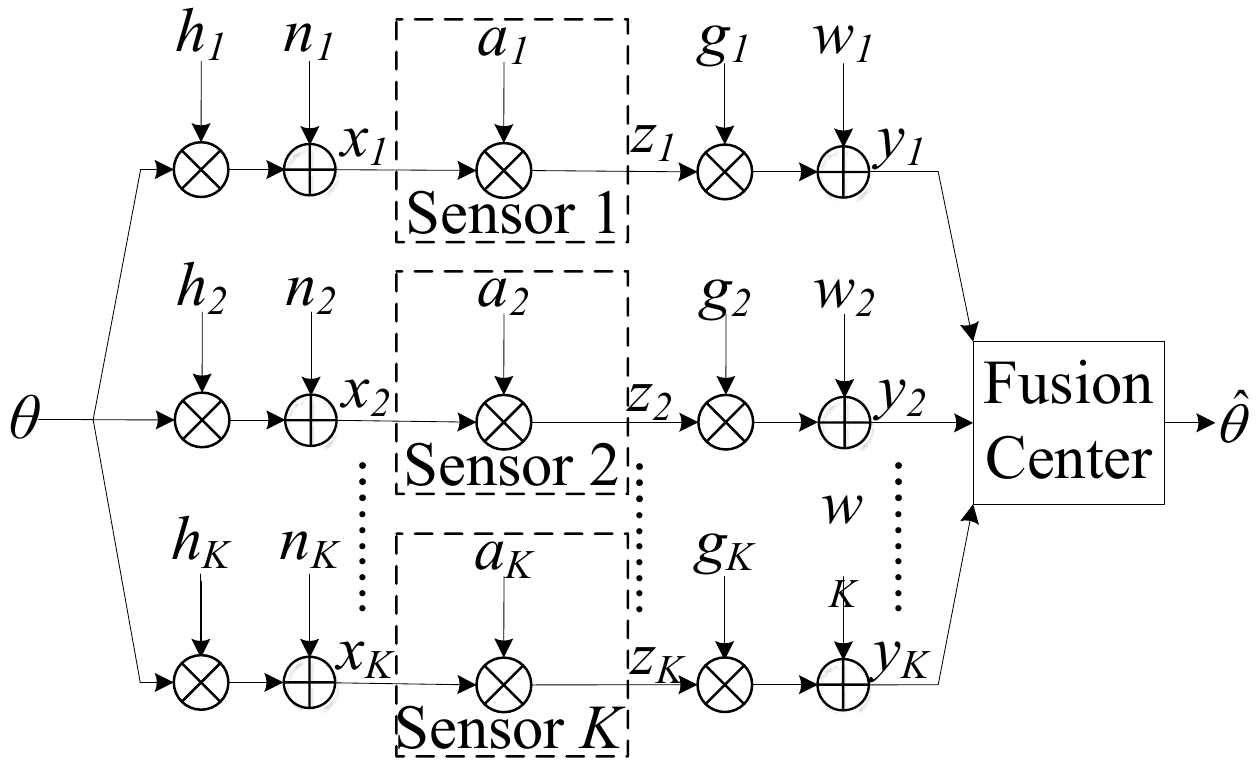}
\caption{System model of a WSN in which the FC finds an estimate of $\theta$.}
%\caption{System model of a WSN composed of $K$ spatially distributed sensors that send linearly amplified versions of their local noisy observations to the FC through parallel channels corrupted by fading and additive Gaussian noise. The FC finds the BLUE estimator of $\theta$.}
\label{Fig:SystemModel}
\end{figure}

Consider a wireless sensor network (WSN) composed of $K$ spatially distributed sensors, as depicted in Fig.~\ref{Fig:SystemModel}. The goal of the WSN is to estimate an unknown random parameter $\theta$ at its fusion center (FC) using amplified versions of local noisy observations received through orthogonal coherent channels corrupted by fading and additive Gaussian noise.
% An example of the unknown parameter to be estimated can be the intensity of the signal broadcast by an energy-emitting source and sensed by a set of locally distributed signal detectors. This variable along with the propagation model of the given signal in the observation environment could then be used to estimate the location of the source.
% temperature of the environment sensed by a set of locally distributed thermometers.
Assume that $\theta$ has zero mean and \ifbool{HomogeneousWSN}{unit power}{variance $\sigma^2_{\theta}$}, and is otherwise unknown.

Suppose that the local noisy observation at each sensor is a linear function of the unknown random parameter as
\be
x_i
& = &
h_i \theta + n_i,
\qquad
i = 1, 2, \dotsc, K,
\ee
where $x_i$ is the local observation at the $i$th sensor, $h_i$ is the fixed local observation gain known at the sensor and FC, and $n_i$ is the spatially independent \ifbool{HomogeneousWSN}{and identically distributed}{} additive observation noise with zero mean and known variance $\SigmaO$. Note that no further assumption is made on the distribution of the random parameter to be estimated and the observation noise. Without loss of generality, we define the {\em observation signal-to-noise ratio} (OSNR) at sensor $i$ as $\beta_i = \frac{\left|  h_i \right|^2 \ifbool{HomogeneousWSN}{}{\sigma^2_{\theta}}}{\SigmaO}$, $i = 1, 2, \dotsc, K$,
%\be 
%\beta_i
%& = &
%\frac{\left|  h_i \right|^2
%	\ifbool{HomogeneousWSN}{}{\sigma^2_{\theta}}
%	}{\SigmaO},
%\qquad
%i = 1, 2, \dotsc, K,
%\ee
where $\left| \cdot \right|$ denotes the absolute-value operation.

We assume that there is no inter-sensor communication and/or collaboration among spatially distributed sensors. Each sensor uses an {\em amplify-and-forward} scheme to amplify its local noisy observation before sending it to the FC as
\be 
z_i
\ = \
a_i x_i
\ = \
a_i h_i \theta + a_i n_i,
\qquad
i = 1, 2, \dotsc, K,
\ee
where $z_i$ is the signal transmitted from sensor $i$ to the FC and $a_i$ is the local amplification gain at sensor $i$. Note that the instantaneous transmit power of sensor $i$ can be found as
\be \label{Eq:PowerDef}
P_i
\ = \
a_i^2 \left( \left|  h_i \right|^2
	\ifbool{HomogeneousWSN}{}{\sigma^2_{\theta}} + \SigmaO \right)
\ = \
a_i^2 \SigmaO \left( 1 + \beta_i \right).
\ee
As it can be seen in \eqref{Eq:PowerDef}, the value of the local amplification gain at each sensor determines the instantaneous transmit power allocated to that sensor. Therefore, we will call any strategy that assigns a set of local amplification gains to sensors a {\em power-allocation scheme}.

All locally processed observations are transmitted to the FC through orthogonal channels corrupted by fading and
additive Gaussian noise. The received signal from sensor $i$ at the FC can be described as
\be 
y_i
& = &
g_i z_i + w_i,
\qquad
i = 1, 2, \dotsc, K,
\ee
where $g_i$ is the multiplicative fading coefficient of the channel between sensor $i$ and the FC, and $w_i$ is the spatially independent \ifbool{HomogeneousWSN}{and identically distributed}{} additive Gaussian channel noise with zero mean and variance $\SigmaC$. In this paper, we assume that the FC can reliably estimate the fading coefficient of the channel between each sensor and itself. This can be achieved by the transmission of pilot sequences from local sensors to FC. Note that in the above model, we have also assumed that each sensor is synchronized with the FC. Without loss of generality, we define the {\em channel signal-to-noise ratio} (CSNR) at sensor $i$ as $\gamma_i = \frac{\left|  g_i \right|^2 }{\SigmaC}$,
$i = 1, 2, \dotsc, K$.
%\be \label{Eq:ChannelSNR}
%\gamma_i
%& = &
%\frac{\left|  g_i \right|^2 }{\SigmaC},
%\qquad
%i = 1, 2, \dotsc, K.
%\ee

%---------------------------------------------------------------------------------------------------------------
\section{Problem Statement}
\label{Sec:ProbStatement}
Suppose that given a power-allocation scheme and a realization of the fading gains, the FC combines the set of received signals from sensors to find the {\em best linear unbiased estimator} (BLUE) for the unknown parameter $\theta$ as~\cite[Chapter 6]{Kay93}
\be
\widehat{\theta}
\ = \
\left(
\sum_{i=1}^K
\frac{h_i^2 a_i^2 g_i^2}{a_i^2 g_i^2 \SigmaO + \SigmaC}
\right)^{-1}
\sum_{i=1}^K
\frac{h_i a_i g_i y_i}{a_i^2 g_i^2 \SigmaO + \SigmaC},
\ee
where the corresponding estimator variance can be found as
%{
%\setlength{\arraycolsep}{0pt}
\be \label{Eq:ThetaVariance}
\text{Var}
\left(
\widehat{\theta}
\big|
\vc{a},\vc{g}
\right)
%& = &
%\left(
%\sum_{i=1}^K
%\frac{h_i^2 a_i^2 g_i^2}{a_i^2 g_i^2 \SigmaO + \SigmaC}
%\right)^{-1} \notag \\
& = &
\ifbool{HomogeneousWSN}{}{\sigma_\theta^2}
\left(
\sum_{i=1}^K
\frac{\beta_i \gamma_i a_i^2 \SigmaO}{1+\gamma_i a_i^2 \SigmaO}
\right)^{-1},
\ee
%}
in which $\vc{a} \triangleq \left[ a_1, a_2, \dotsc, a_K \right]^T$ and $\vc{g} \triangleq \left[ g_1, g_2, \dotsc, g_K \right]^T$ are column vectors containing the set of local amplification gains $a_i$ and fading coefficients of the channels $g_i$, respectively.

Cui~et~al.~\cite{Cui07Diversity} have derived the optimal local amplification gains or equivalently, the optimal power-allocation scheme to minimize the BLUE-estimator variance, defined in~\eqref{Eq:ThetaVariance}, given a total transmission-power constraint in the network as
\be \label{Eq:OptimalGain}
a_i
& = &
\begin{cases}
%	\frac{1}{
%		\ifbool{HomogeneousWSN}{\sigma_{\text{o}}}{\sigma_{\text{o}i}}
%		}
		\sqrt{ \frac{1}{\gamma_i \SigmaO} \left( \sqrt{\delta_i} \, \rho(K_1) - 1 \right) }, & i \leq K_1 \\
	0, & i > K_1
\end{cases},
\ee
where $\delta_i \triangleq \frac{\beta_i \gamma_i}{1 + \beta_i}$, $i = 1, 2, \dotsc, K$,
% is defined as a function of the observation SNR and channel SNR for sensor $i$,
the sensors are sorted
% in the descending order of the values of their $\delta_i$
so that $\delta_1 \geq \delta_2 \geq \cdots \geq \delta_K$, the function $\rho(\cdot)$ is defined for any integer argument $n$ as
\be
\rho(n)
& = &
\frac{P_{\mathsf{Total}} + \sum_{i=1}^n \frac{\beta_i}{\delta_i}}{\sum_{i=1}^n \frac{\beta_i}{\sqrt{\delta_i}}},
\ee
$K_1$ is the {\em unique} largest integer for which $\sqrt{\delta_{K_1}} \, \rho(K_1) > 1$ and $\sqrt{\delta_{K_1+1}} \, \rho(K_1+1) \leq 1$, and $P_{\mathsf{Total}}$ is the constraint on the total power consumed in the entire network so that $\sum_{i=1}^K P_i \leq P_{\mathsf{Total}}$.
%\be 
%\sum_{i=1}^K P_i \leq P_{\mathsf{Total}}.
%\ee
The above power-allocation strategy assigns a zero amplification gain or equivalently, zero transmit power to the sensors for which $\delta_i \leq \left[ \rho(K_1) \right]^{-2}$, because either the sensor's observation SNR or its channel SNR is too low. The assigned instantaneous transmit power to other sensors is non-zero and based on the value of $\delta_i$ for each sensor. Note that based on the above power-allocation scheme, there is a unique one-to-one mapping between $\vc{g}$ and $\vc{a}$ that could be denoted as $\vc{a} = f \left( \vc{g} \right)$.

It can be observed that the optimal power-allocation scheme proposed in~\cite{Cui07Diversity} is based on the assumption that the complete forward channel state information (CSI) is available at local sensors. In other words, Equation~\eqref{Eq:OptimalGain} shows that the optimal value of the local amplification gain at sensor $i$ is a function of its channel SNR $\gamma_i$, which in itself is a function of the instantaneous fading coefficient of the channel between sensor $i$ and the FC.
% (q.v.~Equation~\eqref{Eq:ChannelSNR}).
Therefore, in order for the FC to achieve the estimator variance given by~\eqref{Eq:ThetaVariance}, it must feed the exact instantaneous amplification gain $a_i$ back to each sensor.\footnote{Note that instead of feeding $a_i$ back to each sensor, the FC could send back the fading coefficient of the channel between each sensor and the FC. However, the knowledge of $g_i$ alone is not enough for sensor $i$ to compute the optimal value of its local amplification gain $a_i$. The sensor must also know whether it needs to transmit (i.e., $i \leq K_1$) or stay silent (i.e., $i > K_1$). There are two ways that the extra data can be fed back to the sensors: This information could be encoded in an extra one-bit command instructing the sensor to transmit or stay silent, or the sensor could listen for the entire vector of $\vc{g}$ sent by the FC over a broadcast channel. Sending back each value of $a_i$ avoids the problems of either having to send each sensor an additional bit, or requiring the sensor to listen to the entire vector of $\vc{g}$.} This requirement is not practical in most applications, especially in large-scale WSNs, since the feedback information is typically transmitted through finite-rate {\em digital} feedback links.

%---------------------------------------------------------------------------------------------------------------
%\section{Limited Feedback in Distributed Estimation}
%\label{Sec:LimitedFeedbackIntro}
In this paper, we propose a limited-feedback strategy to alleviate the above-mentioned requirement for infinite-rate digital feedback links from the FC to the local sensors.
% For each channel realization, the FC first finds the optimal power-allocation scheme using the approach proposed in~\cite{Cui07Diversity}. Note that the FC has access to the perfect {\em backward} CSI; i.e., the instantaneous fading gain of the channel between each sensor and itself. Therefore, it can find the {\em exact} power-allocation strategy of the entire network based on~\eqref{Eq:OptimalGain}, given any channel realization. In the next step, the FC broadcasts the {\em index} of the quantized version of the optimized power-allocation vector.
% \footnote{Note that the other potential limited-feedback strategy for this network would be to quantize the vector of instantaneous state information of the channel between each sensor and the FC, so that each sensor can then calculate its optimal local amplification gain based on its received quantized CSI. This approach is beyond the scope of this paper. Similar analysis could be extended to this case.}
In the proposed limited-feedback strategy, the FC and local sensors must agree on a {\em codebook} of the local amplification gains or equivalently, a codebook of possible power-allocation schemes. Before the sensors transmit their amplified observations, the FC reliably estimates the channels between them and itself (i.e., $\vc{g}$), and finds the optimal power allocation to the sensors for the given channel realization, using the approach proposed in~\cite{Cui07Diversity}. Note that the FC has access to the perfect {\em backward} CSI; i.e., the instantaneous fading gain of the channel between each sensor and itself. Therefore, it can find the {\em exact} power-allocation strategy of the entire network based on~\eqref{Eq:OptimalGain}, given any channel realization. The FC will then identify the closest codeword to the optimal power-allocation vector, and broadcasts the {\em index} of that codeword to all sensors over an error-free digital feedback channel.\footnote{Since the rate of the feedback link is very low,
% in the proposed scheme,
an error-free channel can be realized by using capacity-approaching channel codes.} The optimal codebook can be designed offline by quantizing the space of the optimized power-allocation vectors using the {\em generalized Lloyd algorithm}~\cite{GershoGray92} with modified distortion functions.

%---------------------------------------------------------------------------------------------------------------
\section{Codebook Design Using Lloyd Algorithm}
\label{Sec:CodeDesign}
% Suppose that the FC has access to the perfect state information of the channels between local sensors and itself. In practice, this assumption can be satisfied by transmitting training sequences from each sensor to the FC. Furthermore, assume that the digital feedback links between the FC and local sensors are error-free. Since the rate of these links is very low in the proposed limited-feedback strategy, this assumption can reasonably be materialized by using capacity-approaching channel codes.

Let $L$ be the number of feedback bits that the FC uses to quantize the space of the optimal local power-allocation vectors into $2^L$ disjoint regions. Note that $L$ is the total number of feedback bits broadcast by the FC, and {\em not} the number of bits fed back to each sensor.
% Each quantization region is represented by a codeword.
A codeword is chosen in each quantization region. The length of each codeword is $K$, and its $i$th entry is a {\em real} number representing a quantized version of the optimal local amplification gain for sensor $i$. The proposed quantization scheme could then be considered as a mapping from the space of channel state information to a discrete set of $2^L$ length-$K$ real-valued power-allocation vectors.
% Details of this quantization method are as follows.

Let $\mx{C} = \left[ \vc{a}_1 \; \vc{a}_2 \; \cdots \; \vc{a}_{2^L} \right]^T$ be a $2^L \times K$ codebook matrix of the optimal local amplification gains, where $\left[ \mx{C} \right]_{\ell,i}$ denotes its element in row $\ell$ and column $i$ as the optimal gain of sensor $i$ in codeword $\ell$. Note that each $\vc{a}_\ell$, $\ell=1,2,\dotsc, 2^L$ is associated with a realization of the fading coefficients of the channels between local sensors and the FC, denoted by $\vc{g}_\ell$. We apply the generalized Lloyd algorithm with modified distortion functions to solve the problem of vector quantization in the space of the optimal local amplification gains and to design the optimal codebook $\mx{C}$ in an iterative process.
% This algorithm designs the optimal codebook $\mx{C}$ in an iterative process, as explained in the following discussions.

In order to implement the generalized Lloyd algorithm, a distortion metric must be defined for the codebook and for each codeword. Let $D_\text{B} \left( \mx{C} \right)$ denote the average distortion for codebook $\mx{C}$ defined as
\be \label{Eq:BookDistortion}
D_\text{B} \left( \mx{C} \right)
& \triangleq &
\mathbb{E}_{\vc{g}}
\left[
\underset{ \ell \in \left\{ 1,2,\dotsc,2^L \right\} }{\min}
D_\text{W} \left( \vc{a}_\ell | \vc{g} \right)
\right],
\ee
where $\mathbb{E}_{\vc{g}} \left[ \cdot \right]$ denotes the expectation operation with respect to the fading coefficients of the channel and $D_\text{W} \left( \vc{a}_\ell | \vc{g} \right)$ represents the conditional codeword distortion resulting from assigning a suboptimal quantized power-allocation vector $\vc{a}_\ell$ (instead of the optimal one, denoted by $\vc{a}^{\mathsf{OPT}}$ and found using~\eqref{Eq:OptimalGain}), given a realization of the vector of channel fading coefficients $\vc{g}$. We define this conditional codeword distortion as
\be \label{Eq:WordDistortion}
D_\text{W} \left( \vc{a}_\ell | \vc{g} \right)
\; \triangleq \;
\left|
\text{Var}
\left(
\widehat{\theta}
\big|
\vc{a}_\ell,\vc{g}
\right)
-
\text{Var}
\left(
\widehat{\theta}
\big|
\vc{a}^{\mathsf{OPT}},\vc{g}
\right)
\right|,
\ee
where the estimation variance $\text{Var} \left( \widehat{\theta} \big| {}\cdot{}, \vc{g} \right)$ 
% $\text{Var} \left( \widehat{\theta} \big| \vc{a}_\ell,\vc{g} \right)$ and $\text{Var} \left( \widehat{\theta} \big| \vc{a}^{\mathsf{OPT}},\vc{g} \right)$
is found using~\eqref{Eq:ThetaVariance}.
% and $\vc{a}^{\mathsf{OPT}}$ denotes the optimal power-allocation vector associated with the realization of the vector of channel fading coefficients $\vc{g}$. Note that $\vc{a}^{\mathsf{OPT}}$ is found using~\eqref{Eq:OptimalGain}.

Let $\mathcal{G} \subseteq \mathbb{R}^{K+}$ and $\mathcal{A} \subseteq \mathbb{R}^{K+}$ be the $K$-dimensional vector spaces of fading coefficients of the channel and optimal local amplification gains, respectively, whose entries are chosen from the set of real-valued non-negative numbers. Note that each vector of channel fading coefficients $\vc{g} \in \mathcal{G}$ is {\em uniquely} mapped into an optimal power-allocation vector $\vc{a}^{\mathsf{OPT}} \in \mathcal{A}$ by applying~\eqref{Eq:OptimalGain} to find each element of $\vc{a}^{\mathsf{OPT}}$. We denote this mapping by $f: \mathcal{G} \to \mathcal{A}$. Given the distortion functions for the codebook $\mx{C}$ and for each one of its codewords defined in Equations~\eqref{Eq:BookDistortion} and~\eqref{Eq:WordDistortion}, respectively, the two main conditions of the generalized Lloyd algorithm could be reformulated for our vector-quantization problem as follows~\cite[Chapter 11]{GershoGray92}.

%\begin{LaTeXdescription}
%	\item[Nearest--Neighbor Condition:]
	\noindent {\bf Nearest--Neighbor Condition:} This condition finds the optimal quantization regions (or Voronoi cells) for the vector space to be quantized, given a fixed codebook. Based on this condition, each point $\vc{a} \in \mathcal{A}$ in the vector space of the optimal local amplification gains is assigned to partition $\ell$ represented by codeword $\vc{a}_\ell \in \mx{C}$ if and only if its distance to codeword $\vc{a}_\ell$, with respect to the conditional codeword distortion function defined in~\eqref{Eq:WordDistortion}, is less than its distance to any other codeword in the codebook. In this paper, given a codebook $\mx{C}$, the space $\mathcal{G}$ of channel fading coefficients is divided into $2^L$ disjoint quantization regions with the $\ell$th region defined as
	% the space $\mathcal{A}$ of optimized power-allocation vectors is divided into $2^L$ disjoint quantization regions with the $\ell$th region defined as
%\end{LaTeXdescription}
	\be \label{Eq:PartitionDef}
	\mathcal{G}_\ell
	\ = \
	\left\{
	\vc{g} \in \mathcal{G} {}:{}
	D_\text{W} \left( \vc{a}_\ell | \vc{g} \right)
	\leq
	D_\text{W} \left( \vc{a}_k | \vc{g} \right)
	, \forall k \neq \ell
	\right\}.
	\ee
%\begin{LaTeXdescription}
%	\item[\ \ \ \ ]
	Since for every $\vc{g} \in \mathcal{G}$, there is a {\em unique} $\vc{a} \in \mathcal{A}$ that can be found using~\eqref{Eq:OptimalGain}, we could define the Voronoi cell $\mathcal{A}_\ell$ as the image of $\mathcal{G}_\ell$ under mapping $\mathcal{A}_\ell = f \left( \mathcal{G}_\ell \right)$. In other words, to find the optimal partition for each $\vc{a} \in \mathcal{A}$, its corresponding vector of channel fading coefficients $\vc{g} \in \mathcal{G}$ is considered. The distortion of using any codeword $\vc{a}_\ell \in \mx{C}$ instead of the optimal power-allocation vector for that channel realization is found using~\eqref{Eq:WordDistortion}, and $\vc{a}$ is assigned to the region with the lowest conditional codeword distortion, given $\vc{g}$.

%	\item[Centroid Condition:]
	\noindent {\bf Centroid Condition:} This condition finds the optimal codebook, given a specific partitioning of the vector space of the optimized power-allocation vectors $\left\{ \mathcal{A}_1, \mathcal{A}_2, \dotsc, \mathcal{A}_{2^L}  \right\}$. Based on this condition, the optimal codeword associated with each Voronoi cell $\mathcal{A}_\ell \subseteq \mathcal{A}$ is the {\em centroid} of that cell with respect to the conditional codeword-distortion function introduced in~\eqref{Eq:WordDistortion}, and is defined as
	% In this paper, given a specific partitioning of the space of the optimized power-allocation vectors $\left\{ \mathcal{A}_1, \mathcal{A}_2, \dotsc, \mathcal{A}_{2^L}  \right\}$, the optimal codeword associated with partition $\mathcal{A}_\ell$ is defined as
	\be \label{Eq:CodewordDef}
	\vc{a}_\ell
	& = &
	\argmin{\vc{a} \in \mathcal{A}_\ell} \;
	\mathbb{E}_{ \vc{g} \in \mathcal{G}_\ell }
	\left[
	D_\text{W} \left( \vc{a} | \vc{g} \right)
	\right],
	\ee
	where the expectation operation is performed over the set of realizations of the channel fading coefficients, whose associated optimal power-allocation vectors are members of partition $\mathcal{A}_\ell$. This set is denoted by $\mathcal{G}_\ell$.
%\end{LaTeXdescription}

The optimal codebook is designed offline by the FC using the above two conditions. It can be shown that the average codebook distortion is monotonically non-increasing through the iterative usage of the Centroid Condition and the Nearest-Neighbor Condition~\cite[Chapter 11]{GershoGray92}. Details of the codebook design process are summarized in Algorithm~I. The optimal codebook is stored in the FC and all sensors.

\begin{table}[!t]
%	\caption{}
%	\label{Alg:LloydAlg}
	\centering
	\newlength{\MyColumnTextWidth}
	\setlength{\MyColumnTextWidth}{1\linewidth}
	\addtolength{\MyColumnTextWidth}{-1\columnsep}
	\begin{tabular}{m{1\MyColumnTextWidth}}
	\toprule
	\begin{minipage}[c]{1\linewidth}
	\centering
	ALGORITHM I: The process of optimal codebook design based on the generalized Lloyd algorithm with modified distortion functions.
	\end{minipage} \\
	\midrule
	%\begin{minipage}[c]{1\linewidth}
	%\centering
	\alglanguage{pseudocode}
	\renewcommand{\alglinenumber}[1]{{\footnotesize #1}.}
	\begin{algorithmic}[1]
	% \algblock[Init]{Initialization}{EndInit}
	\algsetblock[Init]{Initialization}{EndInitialization}{6}{0cm}
	\Require $K$, $L$, and channel-fading model.
%	\Require Fading model of the channel between local sensors and the FC.
	\Require $M$.\Comment{$M$ is the number of {\em training vectors} in space $\mathcal{A}$.}
	\Require $\epsilon$.\Comment{$\epsilon$ is the distortion threshold to stop the iterations.}
\Initialization
	\State\hspace{\algorithmicindent} \AlgParBox{$\mathcal{G}_s \longleftarrow \ $ A set of $M$ length-$K$ vectors of channel-fading realizations based on the given fading model of the channels between local sensors and the FC.\Comment{$M \gg 2^L$ and $\mathcal{G}_s \subseteq \mathcal{G}$.}}
	% \parbox[t]{\dimexpr\linewidth}{\raggedleft \Comment{$M \gg 2^L$ and $\mathcal{G}_s \subseteq \mathcal{G}$.}\strut}
	\State\hspace{\algorithmicindent} \AlgParBox{$\mathcal{A}_s \longleftarrow \ $ The set of optimal local power-allocation vectors associated with the channel fading vectors in $\mathcal{G}_s$, found by applying Eq.~\eqref{Eq:OptimalGain}.\Comment{$\mathcal{A}_s$ is the set of training vectors, and $\mathcal{A}_s \subseteq \mathcal{A}$.}}
	\State\hspace{\algorithmicindent} \AlgParBox{{\em Randomly} select $2^L$ optimal power-allocation vectors from the set $\mathcal{A}_s$ as the initial set of codewords. Denote the codewords by $\vc{a}_\ell^0$.
		%, $\ell=1,2,\dotsc,2^L$.
		}
	\State\hspace{\algorithmicindent} \AlgParBox{$\mx{C}^0 \longleftarrow \left[ \vc{a}_1^0 \;\; \vc{a}_2^0 \;\; \cdots \;\; \vc{a}_{2^L}^0 \right]^T$\Comment{$\mx{C}^0$ is the initial codebook.}}
	\State\hspace{\algorithmicindent} \AlgParBox{$j \longleftarrow 0$ and $\text{NewCost} \longleftarrow D_\text{B} \left( \mx{C}^0 \right)$}
	\parbox[t]{\dimexpr\linewidth}{\raggedleft \Comment{The average distortion of codebook is found using Eq.~\eqref{Eq:BookDistortion}.}\strut}
%	\State\hspace{\algorithmicindent} \AlgParBox{$j \longleftarrow 0$}
%	\State\hspace{\algorithmicindent} \AlgParBox{$\text{NewCost} \longleftarrow D_\text{B} \left( \mx{C}^0 \right)$}
%	\parbox[t]{\dimexpr\linewidth}{\raggedleft \Comment{The average distortion of codebook is found using Eq.~\eqref{Eq:BookDistortion}.}\strut}
	\EndInitialization
	\Repeat
	%\State \AlgParBox{$j \longleftarrow j+1$}
	%\State \AlgParBox{$\text{OldCost} \longleftarrow \text{NewCost}$}
	\State \AlgParBox{$j \longleftarrow j+1$ and $\text{OldCost} \longleftarrow \text{NewCost}$}
	\State \AlgParBox{Given codebook $\mx{C}^{j-1}$, optimally partition the set $\mathcal{A}_s$ into $2^L$ disjoint subsets based on the {\em Nearest-Neighbor Condition} using Eq.~\eqref{Eq:PartitionDef}. Denote the resulted optimal partitions by $\mathcal{A}_\ell^{j-1}$.
		%, $\ell = 1,2,\dotsc,2^L$.
		}
	% $\mathcal{R} = \left\{ \mathcal{A}_1^j, \mathcal{A}_2^j, \dotsc, \mathcal{A}_{2^L}^j  \right\}$
	\ForAll{$\mathcal{A}_\ell^{j-1}$, $\ell=1,2,\dotsc,2^L$}
	\State \parbox[t]{\dimexpr(\linewidth-\algorithmicindent)-\algorithmicindent}{Find the optimal codeword associated with partition $\mathcal{A}_\ell^{j-1}$ based on the {\em Centroid Condition} using Eq.~\eqref{Eq:CodewordDef}. Denote this new optimal codeword as $\vc{a}_\ell^j$.\strut}
	\EndFor
	\State \AlgParBox{$\mx{C}^j \longleftarrow \left[ \vc{a}_1^j \;\; \vc{a}_2^j \;\; \cdots \;\; \vc{a}_{2^L}^j \right]^T$\Comment{$\mx{C}^j$ is the new codebook.}}
	\State \AlgParBox{$\text{NewCost} \longleftarrow D_\text{B} \left( \mx{C}^j \right)$}
	\Until{$\text{OldCost} - \text{NewCost} \leq \epsilon$}
	%\State $\mx{C}^\mathsf{OPT} \longleftarrow \mx{C}^j$
	%\State \Return $\mx{C}^\mathsf{OPT}$
	\State \Return $\mx{C}^\mathsf{OPT} \longleftarrow \mx{C}^j$
	\end{algorithmic}
	%\end{minipage} 
	\\
	\bottomrule
	\end{tabular}
\end{table}

Upon observing a realization of the channel fading vector $\vc{g}$, the FC finds its associated optimal power-allocation vector $\vc{a}^{\mathsf{OPT}}$. It then identifies the closest codeword in the optimal codebook $\mx{C}$ to $\vc{a}^{\mathsf{OPT}}$ with respect to the conditional codeword distortion defined in~\eqref{Eq:WordDistortion}. Finally, the FC broadcasts the {\em index} of that codeword to all sensors as
\be
\ell
& = &
\argmin{ k \in \left\{ 1,2,\dotsc, 2^L \right\}, \vc{a}_k \in \mx{C} }
D_\text{W} \left( \vc{a}_k | \vc{g} \right).
\ee
Upon reception of the index $\ell$, each sensor $i$ knows its local amplification gain
% or equivalently, its power-allocation weight
as $\left[ \mx{C} \right]_{\ell,i}$, where $\ell$ and $i$ are the row and column indexes of the codebook $\mx{C}$, respectively.

%---------------------------------------------------------------------------------------------------------------
\section{Numerical Results}
\label{Sec:NumResults}
In this section, numerical results are provided to verify the effectiveness of the proposed limited-feedback strategy in achieving a BLUE-estimator variance close to that of a WSN with full feedback. In our simulations,
\ifbool{HomogeneousWSN}{}{we have set $\sigma^2_{\theta} = 1$ and}the local observation gains are randomly chosen from a Gaussian distribution with unit mean and variance 0.09, i.e., $h_i \sim \mathcal{N} \left( 1, 0.09 \right)$. In all simulations, the average power of $h_i$ across all sensors is set to be 1.2. \ifbool{HomogeneousWSN}{The observation and channel noise variances are set to $\SigmaO = 10 \text{ dBm}$ and $\SigmaC = -90 \text{ dBm}$, respectively.}{The observation noise variances $\SigmaO$ are uniformly selected from the interval $(0.05,0.15)$ such that the average power of the noise variances across all sensors is kept at 0.01. The channel noise variance for all sensors is set to $\SigmaC = -90 \text{ dBm}$, $i = 1, 2, \dotsc, K$.} The following fading model is considered for the channels between local sensors and the FC:
\be
g_i
& = &
\eta_0 \left( \frac{d_i}{d_0} \right)^{-\frac{\alpha}{2}} f_i,
\qquad
i = 1, 2, \dotsc, K,
\ee
where $\eta_0 = -30 \text{ dB}$ is the nominal path gain at the reference distance $d_0 = 1$ meter, $d_i$ is the distance between sensor $i$ and the FC, and uniformly distributed between 50 and 150 meters,
% (in meters), $\alpha \geq 2$
$\alpha=2$ is the path-loss exponent, and $f_i$ is the independent and identically distributed (i.i.d.) Rayleigh-fading random variable with unit variance.
% In our simulations, we have set $\alpha=2$.
% The distance between sensors and the FC is uniformly distributed between 50 and 150 meters.
The size of the training set in the optimal codebook-design process described in Algorithm~I is set to $M =$ 5,000. The codebook-distortion threshold for stopping the iterative algorithm is assumed to be $\epsilon = 10^{-6}$. The results are averaged over 50,000 Monte-Carlo simulations.

\begin{figure}[!t]
  \centering
  \subfloat[The number of sensors in the network is $K = 5$.]{\label{Fig:K5_L}\includegraphics[width=0.92\linewidth,page=1]{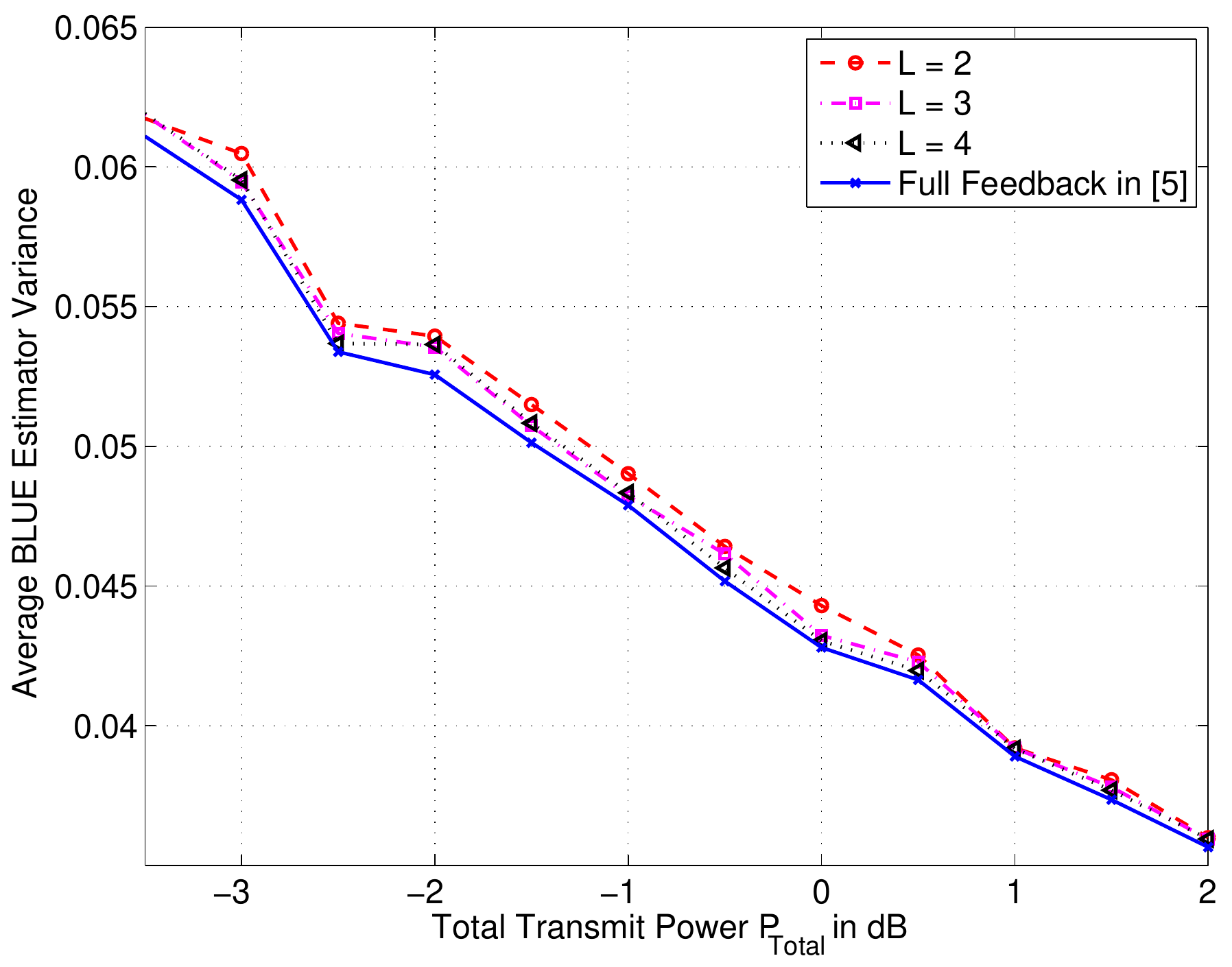}} \\ % \hfill
  ~ %add desired spacing between images, e. g. ~, \quad, \qquad etc. (or a blank line to force the subfig onto a new line)
  %\vspace{0.2cm}
  \subfloat[The number of sensors in the network is $K = 10$.]{\label{Fig:K10_L}\includegraphics[width=0.92\linewidth,page=1]{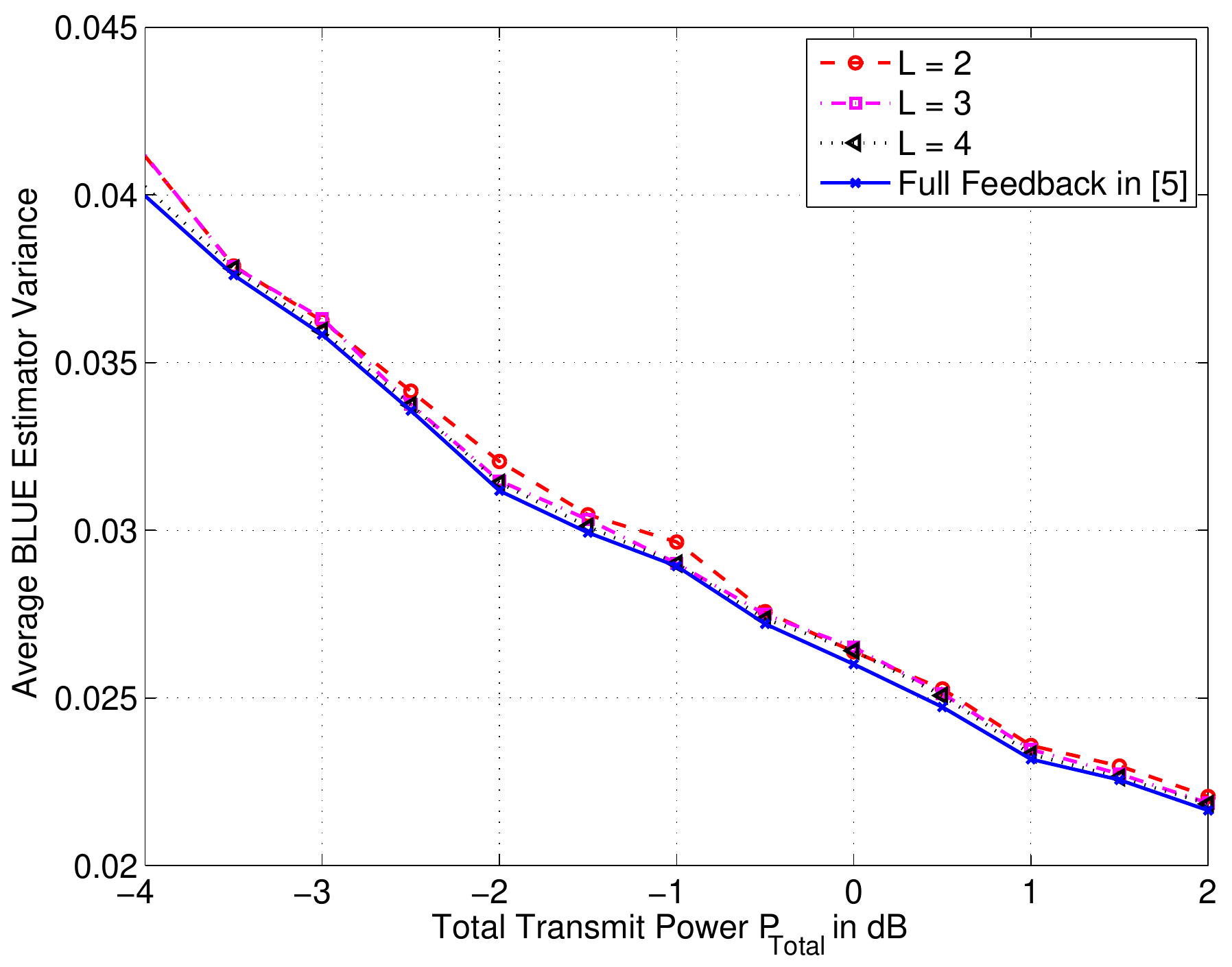}}
  %\vspace{0.2cm}
  \caption{\label{Fig:BlueVar_K} Average BLUE-estimator variance versus the total transmit power $P_{\mathsf{Total}}$ for different values of the number of feedback bits $L$.
  %, when there are~(a)~$K=5$ or~(b)~$K=10$ sensors in the network.
  }
\end{figure}

\begin{figure}[t]
%	\addtolength{\belowcaptionskip}{-10pt}
\centering
\includegraphics[width=0.93\linewidth]{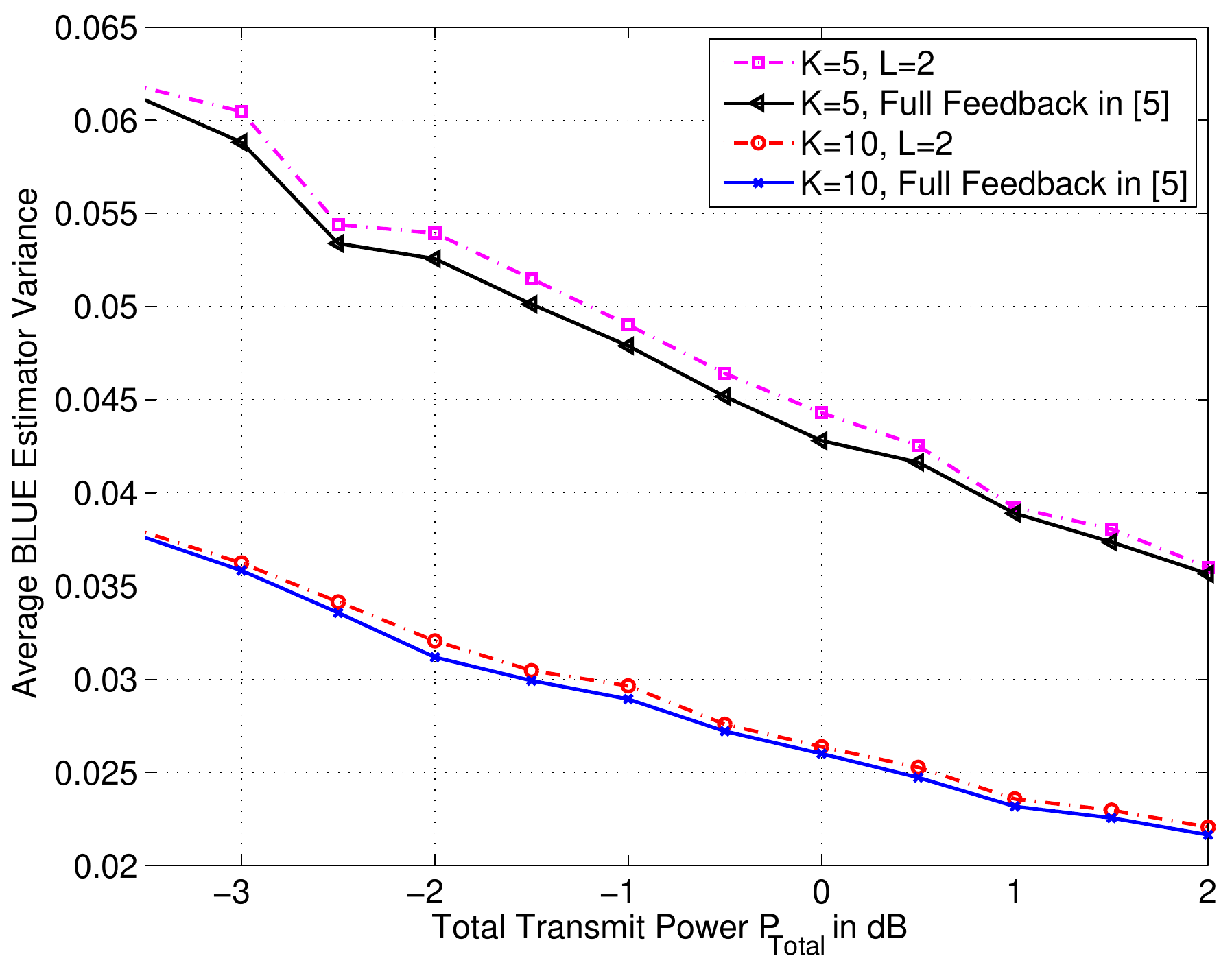}
\caption{Average BLUE-estimator variance versus the total transmit power $P_{\mathsf{Total}}$ for different values of the number of sensors $K$. The number of feedback bits
	%for the limited-feedback case
	is $L=2$.}
\label{Fig:L2_K}
\end{figure}

Figure~\ref{Fig:BlueVar_K} illustrates the effect of $L$ as the number of feedback bits from the FC to local sensors on the performance of the BLUE estimator. It should be emphasized that $L$ is the total number of feedback bits broadcast by the FC, and not the number of bits fed back to each sensor. This figure depicts the average BLUE-estimator variance
%, i.e., $\mathbb{E}_{\vc{g}} \left[ \text{Var} \left( \widehat{\theta} \big| \vc{a},\vc{g} \right) \right]$,
versus the total transmit power $P_{\mathsf{Total}}$ for different values of the number of feedback bits $L$, when there are $K=5$ or $K=10$ sensors in the network, shown in Figs.~\ref{Fig:K5_L} and~\ref{Fig:K10_L}, respectively. The results for the case of full-feedback from the FC to local sensors proposed in~\cite{Cui07Diversity} are shown with solid lines as a benchmark. As it can be seen in this figure, the performance of the BLUE estimator with limited feedback is close to that with full feedback, and gets closer to it as the number of feedback bits is increased from 2 to 4.
% It is important to emphasize that the performance penalty in the average estimator variance for using limited number of feedback bits from the FC to local sensors, even with $L=2$ bits, is significantly small.
In Figure~\ref{Fig:L2_K}, similar results are shown to illustrate the effect of the number of sensors in the network on the performance of the BLUE estimator.
% Again, the results corresponding to the full-feedback case proposed in~\cite{Cui07Diversity} are shown with solid lines.
The number of feedback bits for the limited-feedback case is $L=2$. As expected, the average BLUE-estimator variance decreases substantially as the number of sensors increases.

%---------------------------------------------------------------------------------------------------------------
\section{Conclusions}
\label{Sec:Conclusions}
In this paper, a limited-feedback strategy was proposed to be applied in an adaptive power-allocation scheme for distributed BLUE estimation of a random scalar parameter at the FC of a WSN. The proposed approach eliminates the requirement for infinite-rate feedback of the instantaneous forward CSI from the FC to local sensors in order for them to find their optimal local amplification gains. The generalized Lloyd algorithm with modified distortion functions was used to quantize the vector space of the optimal local amplification gains and to design an optimal codebook for this space. Upon observing the CSI, the FC broadcasts the index of the closest codeword to the corresponding optimal power-allocation vector, rather than feeding back the exact instantaneous CSI. Numerical results showed that even with a small number of feedback bits, the average estimation variance of the BLUE estimator with adaptive power allocation based on the proposed limited-feedback strategy is close to that with perfect CSI feedback.

%---------------------------------------------------------------------------------------------------------------
\appendices
\ifCLASSOPTIONcaptionsoff
  \newpage
\fi

%---------------------------------------------------------------------------------------------------------------
% trigger a \newpage just before the given reference number-used to balance the columns on the last page.
% adjust value as needed - may need to be readjusted if the document is modified later.

% \IEEEtriggeratref{4}
% \pagebreak
% can use a bibliography generated by BibTeX as a .bbl file.
\fontsize{9.4}{12}
\selectfont
\bibliographystyle{IEEEtran}
% argument is your BibTeX string definitions and bibliography database(s).
\bibliography{Refs}

\begin{thebibliography}{1}
\providecommand{\url}[1]{#1}
\csname url@samestyle\endcsname
\providecommand{\newblock}{\relax}
\providecommand{\bibinfo}[2]{#2}
\providecommand{\BIBentrySTDinterwordspacing}{\spaceskip=0pt\relax}
\providecommand{\BIBentryALTinterwordstretchfactor}{4}
\providecommand{\BIBentryALTinterwordspacing}{\spaceskip=\fontdimen2\font plus
\BIBentryALTinterwordstretchfactor\fontdimen3\font minus
  \fontdimen4\font\relax}
\providecommand{\BIBforeignlanguage}[2]{{%
\expandafter\ifx\csname l@#1\endcsname\relax
\typeout{** WARNING: IEEEtran.bst: No hyphenation pattern has been}%
\typeout{** loaded for the language `#1'. Using the pattern for}%
\typeout{** the default language instead.}%
\else
\language=\csname l@#1\endcsname
\fi
#2}}
\providecommand{\BIBdecl}{\relax}
\BIBdecl

\bibitem{Xiao06}
J.-J. Xiao, S.~Cui, Z.-Q. Luo, and A.~J. Goldsmith, ``Power scheduling of
  universal decentralized estimation in sensor networks,'' \emph{IEEE
  Transactions on Signal Processing}, vol.~54, no.~2, pp. 413--422, February
  2006.

\bibitem{Luo05}
Z.-Q. Luo, ``An isotropic universal decentralized estimation scheme for a
  bandwidth constrained ad hoc sensor network,'' \emph{IEEE Journal on Selected
  Areas in Communications}, vol.~23, no.~4, pp. 735--744, April 2005.

\bibitem{Xiao08}
J.-J. Xiao, S.~Cui, Z.-Q. Luo, and A.~J. Goldsmith, ``Linear coherent
  decentralized estimation,'' \emph{IEEE Transactions on Signal Processing},
  vol.~56, no.~2, pp. 757--770, February 2008.

\bibitem{Banavar10}
M.~K. Banavar, C.~Tepedelenlio{\u{g}}lu, and A.~Spanias, ``Estimation over
  fading channels with limited feedback using distributed sensing,'' \emph{IEEE
  Transactions on Signal Processing}, vol.~58, no.~1, pp. 414--425, January
  2010.

\bibitem{Cui07Diversity}
S.~Cui, J.-J. Xiao, A.~J. Goldsmith, Z.-Q. Luo, and H.~V. Poor, ``Estimation
  diversity and energy efficiency in distributed sensing,'' \emph{IEEE
  Transactions on Signal Processing}, vol.~55, no.~9, pp. 4683--4695, Sep.
  2007.

\bibitem{Senol08}
H.~{\c{S}}enol and C.~Tepedelenlio{\u{g}}lu, ``Performance of distributed
  estimation over unknown parallel fading channels,'' \emph{IEEE Transactions
  on Signal Processing}, vol.~56, no.~12, pp. 6057--6068, December 2008.

\bibitem{Kay93}
S.~M. Kay, \emph{Fundamentals of Statistical Signal Processing: Estimation
  Theory}, 1st~ed.\hskip 1em plus 0.5em minus 0.4em\relax New Jersey: Prentice
  Hall, 1993.

\bibitem{GershoGray92}
A.~Gersho and R.~M. Gray, \emph{Vector Quantization and Signal
  Compression}.\hskip 1em plus 0.5em minus 0.4em\relax Boston: Kluwer Academic
  Publishers, 1992.

\end{thebibliography}

\end{document}